\documentclass[12pt]{article}
\pdfoutput=1
\usepackage[utf8]{inputenc}
\usepackage{amsfonts}
\usepackage{float}
\usepackage{bm}
\usepackage{bbm}
\usepackage{natbib}
\usepackage{graphicx}
\graphicspath{ {./art/} }
\usepackage{subfig}
\usepackage{amsthm}
\usepackage{amssymb}
\usepackage{amsmath}
\usepackage{url}
\usepackage{xcolor}
\usepackage[margin=1in]{geometry}
\usepackage{times}
\usepackage{fullpage}
\RequirePackage[colorlinks,citecolor=blue,urlcolor=blue,breaklinks=true]{hyperref}
\usepackage{soul}
\usepackage{booktabs}
\usepackage[ruled,vlined]{algorithm2e}

\newtheorem{theorem}{Theorem}

\newtheorem{lem}[theorem]{Lemma}

\theoremstyle{definition}

\newcommand{\I}{{I}}
\newcommand{\E}{\text{E}}
\newcommand{\pval}{\operatorname{pval}}

\title{Valid inference for regression with best subset selection}
\author{Huiming Lin, Xin Tan, and Meng Li\\
Department of Statistics, Rice University}
\date{}

\begin{document}\sloppy

\maketitle

\begin{abstract}

Best subset selection is widely implemented in statistical software and is routinely used by practitioners in scientific fields for variable selection. However, classical confidence intervals and $p$-values constructed after model selection generally fail to achieve their nominal frequentist guarantees, which can invalidate subsequent findings. In this article, we characterize the altered conditional sampling distributions of pivotal quantities after best subset selection. Building on selective-inference techniques developed in related settings, our finite-sample characterization of the AIC selection event reveals that its geometry is a union of finitely many intervals on the real line. This geometry enables exact conditioning and avoids the excessive conditioning common in other post-selection methods. We use this characterization to develop valid inference procedures that provide $p$-values with nominal Type~I error and confidence intervals with finite-sample coverage guarantees. The proposed methods are easy to implement, computationally efficient, and broadly applicable to other commonly used best subset selection criteria. We also study inference with unknown noise level using a Monte Carlo selective test conditional on the AIC-selected model, which controls finite-sample Type~I error at the nominal level under the selected model null. In an application to a classical U.S. consumption dataset, the proposed confidence intervals lead to different conclusions from the conventional intervals, even when the selected model is the full model, producing interpretable findings that better align with empirical observations.
\end{abstract}


\section{Introduction}

Confidence intervals and $p$-values are fundamental tools for quantifying uncertainty and drawing scientific conclusions. Classical inference procedures assume that the statistical model is specified before the data are observed. In practice, however, models are often built through data-driven procedures to improve efficiency, parsimony, and interpretability. Although such procedures are ubiquitous in medical research, data science, and statistical textbooks~\citep{claeskens2008model,hyndman2018forecasting}, they raise inferential concerns. In particular, valid inference conditional on a selected model differs from valid inference under a fixed, prespecified model; the former is the subject of post-selection inference. Throughout this article, ``valid'' $p$-values and confidence intervals refer to procedures that attain the intended Type~I error rate and frequentist coverage after conditioning on the selected model.

Best subset selection is arguably the most widely used variable-selection method when enumerating all subsets is computationally feasible, and it has attracted renewed attention even for problems with large numbers of covariates~\citep{bertsimas2016best,zhu2020polynomial}. A particularly common criterion is the Akaike information criterion (AIC), which selects the model with the smallest AIC value. Because AIC-based selection is routinely implemented in statistical software, it is important to understand when classical inference after AIC selection is valid and to develop corrected procedures when it is not. We derive the conditional sampling distributions of pivotal quantities after best subset selection. Our finite-sample characterization shows that the best subset selection event has the geometry of a union of finitely many intervals on the real line. This geometry allows us to construct exact conditional $p$-values and confidence intervals that attain the intended Type~I error rate and frequentist coverage. Closest in motivation, \citet{hong2018overfitting} studied undercoverage of classical intervals after AIC-type subset selection from an overfitting and variance-estimation perspective. Our focus is different: we explicitly condition on the AIC selection event and construct post-selection \(p\)-values and confidence intervals.

Recent years have seen substantial development in post-selection inference for different model-selection procedures. \citet{berk2013valid} proposed conservative confidence intervals that are valid uniformly over all selection methods, and \citet{charkhi2018asymptotic} studied \textit{asymptotic} confidence intervals after AIC selection. In this paper, we seek the exact \textit{finite-sample} inference for one particular selection method. Along related lines, \citet{lee2016exact} studied post-selection correction for the lasso at a fixed regularization parameter. \citet{tibshirani2016exact} developed inference for forward stepwise regression, least angle regression, and the lasso path. \citet{loftus2014significance} considered forward stepwise selection with grouped variables, and \citet{markovic2017unifying} studied a broad family of selective-inference problems. We contribute to this literature by providing exact post-selection inference for best subset selection criteria. The resulting selection event has a distinct and transparent geometry. Moreover, our post-AIC confidence intervals do not require the additional conditioning used in many selective-inference methods~\citep{lee2016exact,tibshirani2016exact}. This avoids unnecessary widening caused by over-conditioning.

Post-selection inference with unknown error variance remains challenging, and plug-in variance estimators are commonly used in practice. \citet{tibshirani_rinaldo_tibshirani_wasserman_2018} studied the asymptotic validity of selective inference with unknown variance. We show that although plug-in procedures can approach nominal performance as the sample size increases, they may still inflate Type~I error in small samples. To address this issue, we use a Monte Carlo selective-inference procedure for the unknown-variance setting that yields finite-sample calibration. \citet{fithian_taylor_tibshirani_tibshirani_2015,fithian_sun_taylor_2017} developed post-selection inference for sequential model selection under saturated and selective-model frameworks. Our work extends these frameworks to best subset selection.

The rest of the article is organized as follows. Section~\ref{sec:moti} introduces the post-selection target and presents a motivating example. Section~\ref{sec:method} gives a finite-sample characterization of the AIC selection event and shows that it is a union of finitely many intervals on the real line. This characterization leads to corrected sampling distributions for pivotal quantities and to confidence intervals with guaranteed coverage. Section ~\ref{sec:linear-combinations} discusses the post-selection inference procedure when the parameter is linear combination of coefficients. Section~\ref{selective} presents a Monte Carlo selective framework for best-subset post-selection inference under unknown variance. Section~\ref{sec:simu} reports simulations comparing corrected inference with conventional naive inference for $p$-values and confidence intervals under both known and unknown variance. Section~\ref{sec:application} applies the proposed method to a classical U.S. consumption dataset and shows that post-selection inference can lead to different conclusions from conventional confidence intervals, even when AIC selects the full model. Section~\ref{sec:discuss} concludes with extensions to other information-based criteria.

\section{Post-best-subset inference and a motivating example}\label{sec:moti}
\newcommand{\RSS}{\text{RSS}}
\newcommand{\AIC}{\text{AIC}}
\newcommand{\Shat}{S_0} 
\newcommand{\Sg}{S} 

Consider the linear model (referred to as the full model)
\begin{equation}\label{mdl:lm}
    Y=X\beta+\epsilon,
\end{equation}
where $Y$ is an $n$-vector of responses; $X$ is an $n\times p$ matrix of explanatory variables; $\beta$ is a $p$-vector of coefficients; and $\epsilon\sim N(0,\sigma^2\I)$ is an $n$-vector of independent normal errors. For an index set $S \subset \{1, \ldots, p\}$, we write $X_S$ for the submatrix of $X$ containing the columns indexed by $S$, and $\beta_S$ for the corresponding subvector of $\beta$. We use $|S|$ to denote the cardinality of $S$. We assume that $X$ has full column rank and that $p<n$. Model~\eqref{mdl:lm} is commonly assumed in the model-selection literature as the $\mathcal{M}$-closed case~\citep{Bernardo+Smith:94, Li+Dunson:20}. Under this model, $S^* \subset \{1, \ldots, p\}$ denotes the true model. Our methods, however, apply more generally to any $Y\sim N(\E Y,\sigma^2 I)$ without assumptions on $\E Y$, thereby allowing model~\eqref{mdl:lm} to be misspecified; see Section~\ref{sec:assumption}.

Suppose that a subset $\Shat$ of explanatory variables is first selected by a data-driven procedure and that inference is then based on the selected model. For a target parameter $\eta^T\E Y$ with non-stochastic $\eta\in\mathbb{R}^n$, we aim to characterize the distribution of $\eta^T Y$ conditional on the event that $\Shat$ is selected. For a prespecified null value $c\in\mathbb{R}$, this characterization allows us to construct a valid $p$-value $\pval_c(Y)$ for testing $H_0:\eta^T\E Y=c$ such that
\begin{equation} \label{eq:conditional.p} 
    \Pr(\pval_c(Y) \leq \alpha \mid \Shat \text{ is selected}) \leq \alpha,
\end{equation}
and an interval $(L,U)$ such that
\begin{equation} \label{eq:conditional.coverage} 
    \Pr(\eta^T \E Y \in (L, U) \mid \Shat \text{ is selected}) \geq 1 - \alpha,
\end{equation}
for a given $\alpha\in(0,1)$, where the selection method is best subset selection with AIC. The conditional coverage in~\eqref{eq:conditional.coverage} reflects the shifted goal of post-selection inference relative to the classical setting in which the model is fixed in advance~\citep{berk2013valid}. It also implies a form of control over false positive  discoveries after variable selection~\citep{lee2016exact} and yields unconditional coverage by marginalizing over the event that $\Shat$ is selected~\citep{tibshirani2016exact}.

Under model~\eqref{mdl:lm}, the target parameter $\eta^T\E Y$ equals $\eta^T X\beta$. In the presence of model selection, naively applying classical inference to the selected model may produce miscalibrated coverage because it ignores the randomness in the selection step. In particular, for a selected model $S$, the classical $p$-value for $\eta^T X\beta$ is based on a $t$ distribution, and the usual $1-\alpha$ confidence interval is
\begin{equation}
    \label{eq:CI.t} 
    C(\eta, S, \alpha) = \eta^TX_{S}\hat{\beta}_{S}\pm t_{n-|S|-1,\alpha/2}\hat{\sigma}_{S}\{\eta^TX_{S}(X_{S}^TX_{S})^{-1}X_{S}^T\eta\}^{1/2};
\end{equation}
here $t_{n-|S|-1,\alpha/2}$ is the upper $\alpha/2$ quantile of a $t$ distribution with $n-|S|-1$ degrees of freedom, and $\hat{\beta}_S$ is the least-squares estimate of $\beta_S$ under model $S$. $\hat{\sigma}^2_S=\text{RSS}(S)/(n-|S|-1)$ estimates $\sigma^2$, and $\text{RSS}(S)$ is the residual sum of squares for model $S$. Similarly, when the noise level \(\sigma\) is known, the classical $p$-value uses the normal distribution, and the corresponding interval is $C_{\sigma}(\eta,S,\alpha)
=
\eta^T X_S\hat{\beta}_S
\pm
z_{\alpha/2}\sigma
\{\eta^T X_S(X_S^TX_S)^{-1}X_S^T\eta\}^{1/2},$
where \(z_{\alpha/2}\) is the upper \(\alpha/2\) quantile of the standard normal distribution. 
The classical $p$-value and confidence interval do not, in general, guarantee the conditional Type~I error rate in~\eqref{eq:conditional.p} or the conditional coverage in~\eqref{eq:conditional.coverage}.

We use the following experiment to demonstrate the inflated Type~I error of classical $p$-values and the coverage loss of classical confidence intervals. The data-generating mechanism is similar to that considered by \citet{hong2018overfitting}. We set $n=50$, $p=10$, and $\sigma=1$. The true coefficient vector is $\beta^*=(1,2,3,0,0,\ldots,0)$, with three nonzero components. The rows of $X$ are independently generated from a multivariate Gaussian distribution with mean zero and a first-order autoregressive correlation matrix, AR(1), with correlation parameter $\rho=0.5$. We evaluate the conditional distribution of $p$-values for null coefficients and the coverage probability of $1-\alpha$ confidence intervals for the predicted mean at ten new points $\{x^i\}_{i=1}^{10}$, each generated from the same distribution as the rows of $X$. When $\sigma$ is unknown, the naive $1-\alpha$ confidence interval at point $x$ is $C(x,S,\alpha)$, obtained by substituting $\eta=\eta(x)=X_S(X_S^TX_S)^{-1}x$ in~\eqref{eq:CI.t}.

\begin{figure}[htbp]
  \centering
  \includegraphics[width=0.40\textwidth]{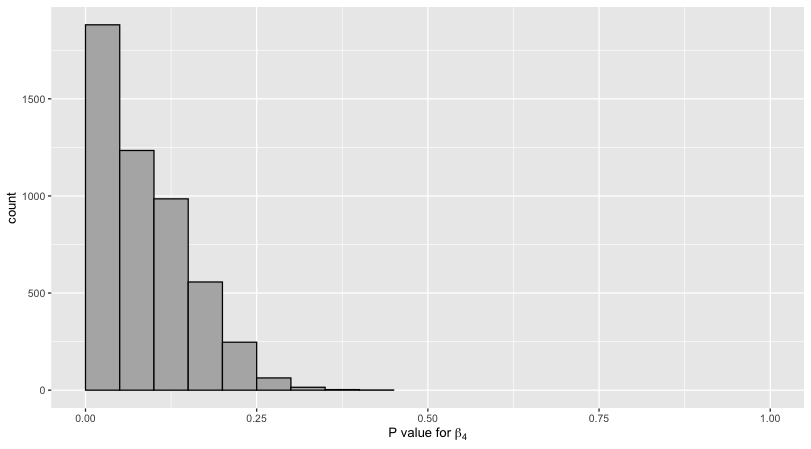}
  \hfill
  \includegraphics[width=0.55\textwidth]{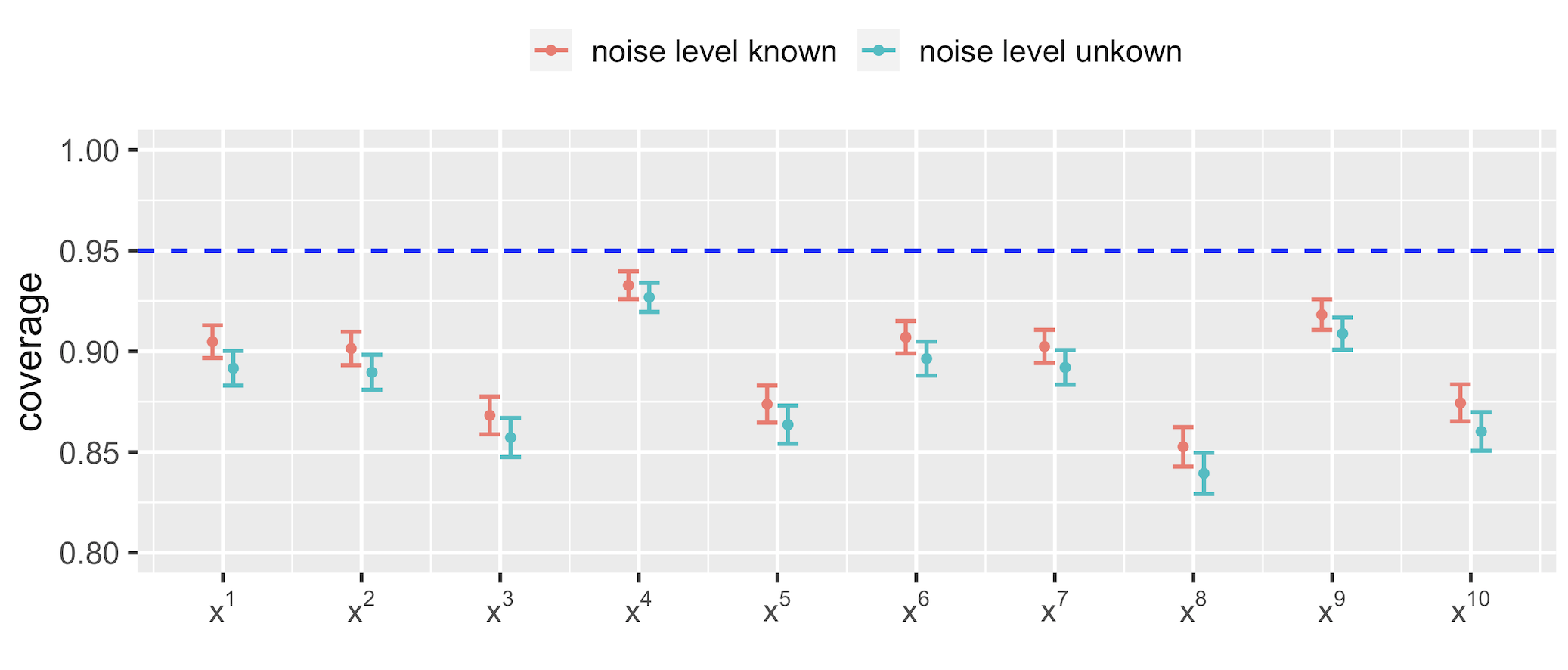}
  \caption{Left: distribution of the conditional $p$-value for $\beta_4$. Right: 95\% confidence intervals for the coverage of $C_{\sigma}(\eta(x),S,0.05)$ (red) and $C(\eta(x),S,0.05)$ (blue) at ten new $x$ points.}
  \label{fig:moti}
\end{figure}

We run 5{,}000 simulations to create a histogram of the $p$-value for $\beta_4$, compute the Type~I error rate, and compute the empirical coverage of the confidence intervals. A common fixed $X$ is used in each simulation, while a new response vector $Y$ is generated in each replication. For each of the ten new data points $\{x^i:1\le i\le 10\}$, we estimate the coverage probability by the frequency with which the interval contains the true mean $(x^i)^T\beta$. Figure~\ref{fig:moti} summarizes the results. The left panel shows the distribution of the $p$-value for $\beta_4$ conditional on its selection. Under $H_0:\beta_4=0$, this distribution is far from uniform; at the nominal level $\alpha=0.05$, the empirical Type~I error rate is $\Pr(p\leq 0.05)=0.377$. The right panel shows undercoverage of classical confidence intervals at the ten new $x$ points under both known and unknown noise levels. These results show that ignoring the selection step can inflate Type~I error in hypothesis testing and reduce the coverage of confidence intervals. In the following sections, we study the distribution of the point estimator $\eta^TY$ conditional on the best subset selection event and show how to conduct valid inference after selection. Section~\ref{sec:simu} shows that the proposed procedures attain nominal Type~I error and confidence-interval coverage.

\section{Statistical inference with post-selection correction}\label{sec:method}
This section develops post-best-subset AIC inference for constructing valid confidence intervals when the same data are used for model selection and model fitting. We begin by characterizing the AIC selection event and showing that it has a simple geometry unique to best subset selection. Extensions to other best subset selection criteria are discussed in Section~\ref{sec:discuss}. The developments in Sections~\ref{sec:AICcharacterize} and~\ref{sec:constructCI} require only $Y\sim N(\E Y,\sigma^2 I)$ and a fixed design matrix $X$; they do not require the linear model assumption in~\eqref{mdl:lm}. This generality is consistent with one premise of AIC, namely that the list of candidate models need not contain the true model.

\subsection{Characterizing the AIC best subset selection event}\label{sec:AICcharacterize}
For the linear regression model~\eqref{mdl:lm}, the AIC of a model $S$ is
\begin{equation}\label{eq:AIC.def}
\text{AIC}(S)=2|S|+n\log\{\text{RSS}(S)\}
\end{equation}
up to an additive constant shared by all models. The selected model $\Shat$ is the model with the smallest AIC value. Let $\mathcal{S}$ denote the set of all candidate submodels, including the full model. Then
\begin{equation}\label{eq:AICselectsShat}
    \{\text{AIC selects  }\Shat\}=\bigcap_{S \in \mathcal{S}:S\neq\Shat}\{\text{AIC}(\Shat)<\text{AIC}(S)\}.
\end{equation}
Thus, characterizing the selection event $\{\text{AIC selects }\Shat\}$ reduces to characterizing the pairwise comparisons $\{\text{AIC}(\Shat)<\text{AIC}(S)\}$ for all $S\neq\Shat$. Let $P_S=\I-X_S(X_S^TX_S)^{-1}X_S^T$ be the residual-maker matrix for model $S$. For a general comparison $\{\text{AIC}(\tilde S)<\text{AIC}(S)\}$, the definition of AIC in~\eqref{eq:AIC.def} gives
\begin{equation}\label{eq:AIC(Shat)<AIC(S)}
    \{\text{AIC}(\tilde{S})<\text{AIC}(S)\}=\left\{\frac{Y^TP_SY}{Y^TP_{\tilde{S}}Y}>\omega(\tilde{S}, S)\right\},
\end{equation}
where $\omega(\tilde{S},S)=\exp\{2(|\tilde{S}|-|S|)/n\}$ for AIC. A representation of~\eqref{eq:AIC(Shat)<AIC(S)} that is more useful for conditional sampling distributions depends on the estimator under study.

Consider a general linear estimator $\eta^TY$ for some non-stochastic $\eta\in\mathbb{R}^n$. The choice of $\eta$ depends on the target parameter and will be specified in later examples. Following standard selective-inference arguments, such as those in \citet{lee2016exact}, we decompose $Y$ into two components: one that is proportional to $\eta^TY$ and one that is independent of it. Let $\tilde{\eta}=\eta(\eta^T\eta)^{-1}$ and
\[
z = Y-(\eta^TY)\tilde{\eta}=\{\I-\eta(\eta^T\eta)^{-1}\eta^T\}Y =: P_{\eta^\perp}Y.
\]
Then $z$ and $\eta^TY$ are uncorrelated, and hence independent under normality. Thus $Y=(\eta^TY)\tilde{\eta}+z$. Substituting this decomposition into~\eqref{eq:AIC(Shat)<AIC(S)} gives, for a fixed $\tilde S$,
\begin{equation}\label{eq:AIC(Shat)<AIC(S)-2}
\begin{split}
    \{\text{AIC}(\tilde{S})&<\text{AIC}(S)\}=\{[\tilde{\eta}^T\{P_S-\omega(\tilde{S}, S)P_{\tilde{S}}\}\tilde{\eta}]\cdot(\eta^TY)^2\\ &+2[z^T\{P_S-\omega(\tilde{S}, S)P_{\tilde{S}}\}\tilde{\eta}]\cdot(\eta^TY)
    +z^T\{P_S-\omega(\tilde{S}, S)P_{\tilde{S}}\}z>0\}.
\end{split}
\end{equation}
The right-hand side of~\eqref{eq:AIC(Shat)<AIC(S)-2} is a quadratic inequality in $\eta^TY$ when the leading coefficient is nonzero. Depending on the sign of $\tilde{\eta}^T\{P_S-\omega(\tilde{S}, S)P_{\tilde{S}}\}\tilde{\eta}$, the feasible region for $\eta^TY$ is either one interval or the union of two intervals. When the leading coefficient is zero, the inequality is linear and the feasible region is a single interval. Hence the comparison can be represented in the following unified form:
\begin{equation} \label{eq:AIC.individual}
    \{\text{AIC}(\tilde{S})<\text{AIC}(S)\} =  \{\eta^T Y \in (a_{1}(z|\tilde{S}, S), b_{1}(z|\tilde{S}, S))\cup(a_{2}(z|\tilde{S}, S), b_{2}(z|\tilde{S}, S))\},
\end{equation}
where some endpoints may be infinite.

Combining~\eqref{eq:AIC.individual} and~\eqref{eq:AICselectsShat} yields the characterization
\begin{equation} \label{eq:selection.event} 
\{\text{AIC selects }\Shat\} = \left\{\eta^TY \in \text{Re}(z | \Shat) \right\},
\end{equation}
where
\begin{equation}
    \label{eq:AIC.selection.event} 
    \text{Re}(z | \Shat) = \bigcap_{S:S\neq\Shat}\{(a_{1}(z|\Shat, S), b_{1}(z|\Shat, S))\cup(a_{2}(z|\Shat, S), b_{2}(z|\Shat, S))\}
\end{equation}
is a union of finitely many intervals. This geometry has explicit analytic forms and differs from the polyhedral conditioning sets obtained for the lasso by \citet{lee2016exact}. Unlike procedures that additionally condition on selected signs~\citep{lee2016exact} or on other enlarged events~\citep{tibshirani2016exact}, our AIC characterization yields exact conditioning. As a result, it can avoid the unnecessary widening of intervals caused by extra conditioning. Section~\ref{sec:simplification} gives an additional simplification for post-AIC inference.

For notational simplicity, and with the selected model understood to be fixed by conditioning, we write $\text{Re}(z)$ for $\text{Re}(z|\Shat)$ below. This convention is analogous to the common practice of suppressing the conditioning on the design matrix $X$ in fixed-design linear models. We also write $\omega(S)$ for $\omega(S_0,S)$.

\subsection{Post-AIC best subset selection inference with known noise level}\label{sec:constructCI}

We now construct $p$-values and confidence intervals for the parameter $\eta^T\E Y$ when $\sigma$ is known. Since $\eta^TY$ is an unbiased estimator of $\eta^T\E Y$, the characterization above gives its conditional distribution after selection. For any $z_0\in\mathbb{R}$,
\renewcommand{\Re}{\text{Re}}
\begin{equation}\label{eq:aic}
\eta^TY|\{\text{AIC selects }\Shat, z=z_0\} = \eta^TY|\left\{\eta^TY \in \Re(z), z = z_0\right\} \overset{d}{=}\eta^TY|\{\eta^TY \in \text{Re}(z_0)\},
\end{equation}
where the last equality uses the independence of $z$ and $\eta^TY$. The conditional distribution in~\eqref{eq:aic} is therefore a truncated normal distribution. This departure from the normal distribution assumed by classical inference is the source of the needed post-selection correction.

Let $F_{\mu,\lambda,R}$ denote the cumulative distribution function (cdf) of a normal random variable with mean $\mu$ and standard deviation $\lambda$, truncated to a region $R\subset\mathbb{R}$ that is a union of finitely many intervals:
$$
F_{\mu,\lambda,R}(x) = \Phi((-\infty, x] \cap R; \mu, \lambda) / \Phi(R; \mu, \lambda),
$$
where $\Phi(\cdot;\mu,\lambda)$ is the probability measure of a $N(\mu,\lambda^2)$ random variable. Under the normal assumption on $Y$, $\eta^TY\sim N(\eta^T\E Y,\sigma^2\eta^T\eta)$. Hence, by~\eqref{eq:aic}, for any $z_0\in\mathbb{R}$, the cdf of $\eta^TY|\{\eta^TY\in\Re(z),z=z_0\}$ is $F_{\eta^T\E Y,\sigma\sqrt{\eta^T\eta},\text{Re}(z_0)}(\cdot)$. Applying the probability integral transform yields
\begin{align*}
& F_{\eta^T\E Y, \sigma\sqrt{\eta^T\eta},\text{Re}(z)}(\eta^TY)|\{\eta^TY\in\text{Re}(z), z=z_0\} \\
\overset{d}{=}{}& F_{\eta^T\E Y, \sigma\sqrt{\eta^T\eta},\text{Re}(z_0)}(\eta^TY)|\{\eta^TY\in\text{Re}(z_0)\} \sim \text{Unif}(0,1),
\end{align*}
where the second relation again uses the independence of $\eta^TY$ and $z$. Thus, conditional on $\{\eta^TY\in\text{Re}(z)\}$, the distribution of $F_{\eta^T\E Y,\sigma\sqrt{\eta^T\eta},\text{Re}(z)}(\eta^TY)$ given $z=z_0$ does not depend on $z_0$. Integrating over $z$ gives
\begin{equation*}
 F_{\eta^T\E Y, \sigma\sqrt{\eta^T\eta},\text{Re}(z)}(\eta^TY)|\{\eta^TY\in\text{Re}(z)\} \sim \text{Unif}(0,1).
\end{equation*}
Combining this display with~\eqref{eq:selection.event}, we obtain
\begin{equation}\label{eq:unif}
 F_{\eta^T\E Y, \sigma\sqrt{\eta^T\eta},\text{Re}(z)}(\eta^TY)|\{\text{AIC selects }\Shat\} \sim \text{Unif}(0,1).
\end{equation}
This pivot yields post-selection $p$-values. For testing $H_0:\eta^\top\E Y=c$ against $H_1:\eta^\top\E Y\neq c$, conditional on the event that AIC selects $S_0$, the two-sided $p$-value is
$$
\pval_c(Y) =  2\min\{F_{c,\sigma\sqrt{\eta^\top\eta},\mathrm{Re}(z)}(\eta^\top Y),
1-F_{c,\sigma\sqrt{\eta^\top\eta},\mathrm{Re}(z)}(\eta^\top Y)\}.
$$
A conditional $1-\alpha$ confidence region for $\eta^T\E Y$ is
$$
\{t:\alpha/2\le F_{t,\sigma\sqrt{\eta^T\eta},\text{Re}(z)}(\eta^TY)\le 1-\alpha/2\}.
$$
Because the normal distribution has a monotone likelihood ratio in its mean parameter, the truncated-normal cdf is monotone decreasing in the mean parameter as well; see Lemma A.1 of \citet{lee2016exact}. Therefore, the confidence region is an interval $(L,U)$ satisfying
\begin{equation}\label{eq:cCI_knownSigma}
    F_{L,\sigma\sqrt{\eta^T\eta},\text{Re}(z)}(\eta^TY)=1-\alpha/2,\quad
    F_{U,\sigma\sqrt{\eta^T\eta},\text{Re}(z)}(\eta^TY)=\alpha/2.
\end{equation}
The construction applies to any estimator of the form $\eta^TY$. A confidence interval $(L,U)$ satisfying~\eqref{eq:cCI_knownSigma} is a valid $1-\alpha$ confidence interval for $\eta^T\E Y$. Section~\ref{sec:linear-combinations} applies this construction to common targets, including the mean response at a new point and individual regression coefficients, and derives useful simplifications.

\subsection{Post-AIC best subset selection inference with unknown noise level}\label{sec:unknown}

In the known-variance case, the truncation region for $\eta^TY$ depends on the observed $z_0$, while $\eta^TY$ remains independent of $z_0$. This independence yields the truncated Gaussian pivot above. When $\sigma$ is unknown, however, the plug-in statistic $T=\eta^TY/(\hat\sigma\|\eta\|_2)$ is no longer pivotal: the distribution of $\hat\sigma$ depends on both $Y$ and the selection event, and $\eta^TY$ and $\hat\sigma$ are no longer independent after conditioning on selection. Consequently, $T$ does not have a truncated $t$ distribution or another simple analytic form. One way to remove the nuisance scale is to condition on the norm $\|Y\|$ in addition to $z=P_{\eta^\perp}Y$ and the selection event. The resulting inference target is $\eta^T\E Y$ conditional on $\{z=z_0,\mathrm{AIC},\|Y\|^2\}$. As noted by \citet{fithian_taylor_tibshirani_tibshirani_2015,fithian_sun_taylor_2017}, this extra conditioning causes the support of $\eta^T\E Y$ to collapse to at most two points through $\|Y\|^2=\|z_0\|^2+(\eta^T\E Y)^2/(\eta^\top\eta)$. This severe loss of information makes the saturated test with unknown $\sigma$ impractical.

A common practical compromise is to estimate $\sigma$ from the selected model and plug the estimate into the truncated Gaussian pivot, treating $\sigma$ as known. This plug-in procedure is asymptotically valid under conditions studied by \citet{tibshirani_rinaldo_tibshirani_wasserman_2018}. In small samples, however, it can still inflate Type~I error at true nulls because it does not fully account for the uncertainty introduced by model selection and variance estimation. Section~\ref{selective} introduces an alternative selective-model inference procedure that avoids conditioning on $z_0$ and attains nominal Type~I error in finite samples.

\newcommand{\Col}{\text{Col}} 
\section{Applications to linear combinations of coefficients} \label{sec:linear-combinations}
\subsection{Parameters of interest and model assumptions} \label{sec:assumption}

We study the widely used plug-in least-squares estimator $c_{\Shat}^T\hat\beta_{\Shat}$, where $\hat\beta_{\Shat}=(X_{\Shat}^TX_{\Shat})^{-1}X_{\Shat}^TY$ is the least-squares estimator in the selected model. This estimator can be written as $c_{\Shat}^T\hat\beta_{\Shat}=\eta^TY$, with $\eta=X_{\Shat}(X_{\Shat}^TX_{\Shat})^{-1}c_{\Shat}$, which depends only on $X_{\Shat}$ and $c_{\Shat}$. Therefore, \eqref{eq:cCI_knownSigma} can be used to construct confidence intervals for $c_{\Shat}^T\E\hat\beta_{\Shat}$. We next clarify the class of target parameters covered by this construction.

For a given model $S$, such as the selected model $\Shat$, define
\begin{equation}\label{eq:parameter.adj}
\beta^{S} = (X_S^T X_S)^{-1} X_S^T \E Y = \arg\min_b \E \| Y - X_S b\|^2
\end{equation}
as the adjusted regression coefficient in model $S$, following \citet{lee2016exact}.  We conduct inference on this target while leaving $\E Y$ unrestricted, thereby accommodating model misspecification; in particular, we do not require $\E Y\in\operatorname{col}(X_S)$. This is in line with the saturated-model framework of \cite{tibshirani2016exact}. Any linear combination
$c_S^\top\beta^S$ can be written as
\[
c_S^\top\beta^S=\eta_S^\top\E Y,
\qquad
\eta_S=X_S(X_S^\top X_S)^{-1}c_S.
\]
Thus, Equation~\eqref{eq:cCI_knownSigma} can be used to construct a confidence interval for $c_{\Shat}^\top\beta^{\Shat}$. Section~\ref{selective} introduces the alternative selected-model framework and compares its conceptual and empirical behavior with that of the saturated-model framework.

Under model~\eqref{mdl:lm}, the parameter $\beta^S$ defined in~\eqref{eq:parameter.adj} is model dependent and generally varies from model to model. To see this, model~\eqref{mdl:lm} implies \(\E Y=X\beta\), and hence
$\beta^S=(X_S^TX_S)^{-1}X_S^TX\beta.$ Therefore, the population coefficient associated with the same regressor may change as \(S\) varies. This creates a dual dependence of the target parameter on \(S\): the identity of the target changes when the selected variables change, and the true value attached to a given target may also vary through the population projection. \citet{tibshirani2016exact} characterized this parameter as a \textit{moving target}. We refer to it as a ``doubly moving target'' to emphasize that both the identity of the target and its value can move with the selected model.

\subsection{Simplified implementation for post-AIC correction} \label{sec:simplification}
For the parameter $c_{\Shat}^T\beta_{\Shat}$, the estimator $c_{\Shat}^T\hat\beta_{\Shat}=\eta^TY$, with $\eta=X_{\Shat}(X_{\Shat}^TX_{\Shat})^{-1}c_{\Shat}$, can be used in~\eqref{eq:cCI_knownSigma}. Two simplifications are useful: an \textit{algebraic} simplification of the interval endpoints and a \textit{distributional} simplification that reduces computation.

The algebraic simplification follows because $\eta$ lies in the column space of $X_{\Shat}$. Hence $P_{\Shat}\tilde\eta=0$, $\tilde\eta^TP_{\Shat}\tilde\eta=0$, and $z^TP_{\Shat}\tilde\eta=0$. Substituting these identities into~\eqref{eq:AIC(Shat)<AIC(S)-2} gives
\begin{equation}\label{eq:simplePairwise}
    \{\text{AIC}(\Shat)<\text{AIC}(S)\}=\{(\tilde{\eta}^TP_S\tilde{\eta})(\eta^TY)^2+2(z^TP_S\tilde{\eta})(\eta^TY)
    +z^T\{P_S-\omega(S)P_{\Shat}\}z>0\}.
\end{equation}
Because $P_S$ is idempotent, the leading coefficient is positive unless $P_S\tilde\eta=0$. We first consider the case with positive leading coefficient; the second simplification below covers the zero case. If the discriminant $(z^TP_S\tilde\eta)^2-\tilde\eta^TP_S\tilde\eta\,z^T\{P_S-\omega(S)P_{\Shat}\}z$ is negative, then $\{\text{AIC}(\Shat)<\text{AIC}(S)\}=(-\infty,\infty)$ and the comparison can be ignored. Equivalently, one may set $a_1(z|\Shat,S)=a_2(z|\Shat,S)=-\infty$ and $b_1(z|\Shat,S)=b_2(z|\Shat,S)=\infty$. If the discriminant is nonnegative, the endpoints in~\eqref{eq:AIC.selection.event} simplify to
\begin{equation} \label{eq:ab.set}
\begin{split}
    b_1(z|\Shat, S)&=\frac{-z^TP_S\tilde{\eta}-\sqrt{(z^TP_S\tilde{\eta})^2-\tilde{\eta}^TP_S\tilde{\eta}z^T\{P_S-\omega(S)P_{\Shat}\}z})}{\tilde{\eta}^TP_S\tilde{\eta}},\\
    a_2(z|\Shat, S)&=\frac{-z^TP_S\tilde{\eta}+\sqrt{(z^TP_S\tilde{\eta})^2-\tilde{\eta}^TP_S\tilde{\eta}z^T\{P_S-\omega(S)P_{\Shat}\}z})}{\tilde{\eta}^TP_S\tilde{\eta}},\\
    a_1(z|\Shat, S)&=-\infty,\quad b_2(z|\Shat, S)=\infty.
\end{split}
\end{equation}

The second simplification concerns pairwise comparisons between $\Shat$ and its supersets. Although these comparisons are part of the selection event, they do not affect the conditional distribution of $\eta^TY$.
\begin{lem}\label{lem:aiceffect_largeS}
Suppose $\Shat\subset S$. Then the event $\text{AIC}(\Shat)<\text{AIC}(S)$ is equivalent to $z^TP_Sz-\omega(S)z^TP_{\Shat}z>0$. Consequently, this event is independent of $\eta^TY$.
\end{lem}
\begin{proof}
Since $\eta$ lies in the column space of $X_{\Shat}$, it also lies in the column space of $X_S$ when $\Shat\subset S$. Hence $P_S\tilde\eta=0$, which implies $\tilde\eta^TP_S\tilde\eta=0$ and $z^TP_S\tilde\eta=0$. Substituting these identities into~\eqref{eq:simplePairwise} gives the stated equivalence. Independence follows from the independence of $\eta^TY$ and $z$.
\end{proof}
Lemma~\ref{lem:aiceffect_largeS} shows that comparisons $\{\text{AIC}(\Shat)<\text{AIC}(S):S\supset\Shat\}$ have no effect on the conditional distribution of $\eta^TY$ and hence no effect on the conditional coverage of intervals based on $\eta^TY$. Therefore, all $2^{p-|\Shat|}-1$ comparisons between $\Shat$ and its supersets can be skipped when constructing valid post-AIC confidence intervals for any $\eta$ derived from the selected model. This less-than-exact conditioning substantially reduces computation, especially when $\Shat$ is small relative to the full model.

\subsection{Special examples}
We consider two common linear combinations of regression coefficients:
\begin{description}
    \item[Example 1] Revisit the motivating example in Section~\ref{sec:moti}, where the target is the response mean at a new data point $x$, namely $x^T\beta=x_{\Shat}^T\beta_{\Shat}$. The plug-in least-squares estimator is $x_{\Shat}^T\hat\beta_{\Shat}=\eta^TY$, with $\eta=X_{\Shat}(X_{\Shat}^TX_{\Shat})^{-1}x_{\Shat}$.
    \item[Example 2] Consider the coefficient $\beta_i$ for any $i\in\Shat\subset\{1,2,\ldots,p\}$. We focus only on coefficients included in the selected model. Let $e_i=:e_i^{\Shat}$ denote the unit vector whose element corresponding to $\beta_i$ is one and whose other elements are zero, so that $\beta_i=e_i^T\beta_{\Shat}$. The estimator is $\eta^TY:=\hat\beta_i=e_i^T(X_{\Shat}^TX_{\Shat})^{-1}X_{\Shat}^TY$, with $\eta=X_{\Shat}(X_{\Shat}^TX_{\Shat})^{-1}e_i$.
\end{description}

For both examples, \eqref{eq:cCI_knownSigma}, together with the simplifications in Section~\ref{sec:simplification}, characterizes the distribution of $\eta^TY$ after best subset selection. This yields valid $p$-values and confidence intervals for the target parameter. The same characterization also explains why classical confidence intervals lose coverage. Consider Example 2. A classical $(1-\alpha)$ confidence interval for $\beta_i$ is obtained by inverting an event of the form $(\eta^TY-\beta_i)^2\le c_\alpha$, whose probability is $1-\alpha$ when the model is fixed in advance. The pivot $\eta^TY-\beta_i$ is centered normal under the classical regime. The next lemma shows that, when $\beta_i$ is a true null, a subset of the pairwise AIC comparisons removes the highest-density region of this normal distribution. This explains why classical inference can produce inflated Type~I error and false discoveries after selection.

Lemma~\ref{lem:aiceffect_largeS} considered comparisons with $S\supset\Shat$. We now consider the case $S\subset\Shat$.
\begin{lem}\label{lem:aiceffect_smallS}
Suppose $S\subset\Shat$ and $i \in \Shat$. If model $S$ does not contain $\beta_i$, then the event $\text{AIC}(\Shat)<\text{AIC}(S)$ is equivalent to
\begin{equation} \label{eq:example.2}
    \frac{1}{\lVert\eta\rVert_2^2}(\eta^TY)^2>\omega(S) z^TP_{\Shat}z-z^TP_{S}z.
\end{equation}
Consequently, a superset of the event $\{\text{AIC selects }\Shat\}$ is
\begin{equation} \label{eq:lower.bound}
     \bigcap_{S: S \subset \Shat\backslash\{i\}} \{\text{AIC}(\Shat)<\text{AIC}(S)\} = \left\{(\eta^TY)^2 > {\lVert\eta\rVert_2^2} \max_{S \subset \Shat\backslash\{i\}} \{\omega(S) z^TP_{\Shat}z-z^TP_{S}z \}
    \right\}.
\end{equation}
\end{lem}

\begin{proof}
It suffices to show~\eqref{eq:example.2}. Since $S\subset\Shat$, there exists a matrix $A_S$ with entries in $\{0,1\}$ that extracts the corresponding rows of $X_{\Shat}^T$; that is, $X_S^T=A_SX_{\Shat}^T$. If $S$ does not contain $i$, then $A_Se_i=0$. Therefore,
\[
X_S^T\eta=X_S^TX_{\Shat}(X_{\Shat}^TX_{\Shat})^{-1}e_i
=A_SX_{\Shat}^TX_{\Shat}(X_{\Shat}^TX_{\Shat})^{-1}e_i=A_Se_i=0.
\]
It follows that $X_S(X_S^TX_S)^{-1}X_S^T\eta=0$ and hence $P_S\eta=\eta$. Thus
\begin{equation*}
\begin{split}
    \tilde{\eta}^TP_S\tilde{\eta}&=(\eta^T\eta)^{-2}\eta^TP_S\eta=
            \frac{1}{\lVert\eta\rVert_2^2},\\
    z^TP_S\tilde{\eta}&=z^T\eta(\eta^T\eta)^{-1}=0.
\end{split}
\end{equation*}
Substitution into~\eqref{eq:simplePairwise} completes the proof.
\end{proof}

Lemma~\ref{lem:aiceffect_smallS} shows that, when $S\subset\Shat$ and $S$ does not contain $i$, the comparison $\{\text{AIC}(\Shat)<\text{AIC}(S)\}$ places a lower bound on $|\eta^TY|$. If $\beta_i$ is a true null, so that $\beta_i=0$, then the right-hand side of~\eqref{eq:lower.bound} can be written as
\begin{equation}
     (\eta^TY - \beta_i)^2 > {\lVert\eta\rVert_2^2} \max_{S \subset \Shat\backslash\{i\}} \{\omega(S) z^TP_{\Shat}z-z^TP_{S}z \},
\end{equation}
which removes a high-density region from a centered normal distribution. This tends to reduce the coverage probability of classical confidence intervals and increases false discoveries relative to a procedure with valid post-selection correction. Figure~\ref{fig:pivot} revisits the setting of Section~\ref{sec:moti} and compares the pivotal quantities for two falsely selected variables, $\beta_4$ and $\beta_9$, whose true coefficients are zero. The truncated support of the adjusted distributions highlights the effect of conditioning on variable selection. This effect is especially relevant when the selected model is the full model, because all other candidate models are then nested within the selected model; in this case, classical Type~I error can be inflated and post-selection correction is particularly important.

\section{Monte Carlo selective inference}
\label{selective}
The preceding sections developed post-best-subset inference without assuming that the selected model is correctly specified. In this framework, which \cite{fithian_sun_taylor_2017} refers to as saturated-model, $\E Y\in\mathbb{R}^n$ is unrestricted, and the target is the least-squares functional $\theta=\eta^T\E Y$, regardless of possible model misspecification. The null hypothesis is
$$
H_0^{\mathrm{sat}}:\quad \eta^T\E Y=c, \qquad \E Y\in\mathbb{R}^n.
$$
To eliminate the nuisance components orthogonal to $\eta$, the construction conditions on $P_{\eta^\perp}Y$ and on the selection event. When $\sigma$ is known, this yields the truncated normal pivot for $\eta^TY$ derived in Section~\ref{sec:constructCI}. When $\sigma$ is unknown, a plug-in procedure is convenient, but it can inflate Type I error in small samples, as discussed in Section~\ref{sec:unknown}. 

In this section, we instead use the selective-model selective-testing framework of \citet{fithian_taylor_tibshirani_tibshirani_2015,fithian_sun_taylor_2017} in the AIC best-subset setting. Let $Y_{\mathrm{obs}}$ denote the observed response vector, and let
$
\Shat=\mathcal A_{\AIC}(Y_{\mathrm{obs}},X)
$
be the model selected by AIC, where $\mathcal A_{\AIC}(\cdot,X)$ denotes the AIC best-subset selection map. In the selective-model framework, the selected model is treated as the working model after conditioning on the selection event. The corresponding null is
$$
H_0^{\mathrm{sel}}:\quad \eta^T\E Y=c, \qquad \E Y\in \mathrm{span}(X_{\Shat}).
$$
Thus, unlike the saturated null, the selective-model null imposes a structural assumption on the mean vector after selection. With a stronger modeling assumption, this procedure conditions on less information and can retain more information for testing.

For a coefficient contrast in the selected model, write the target as $a^T\beta^{\Shat}$, where $a\in\mathbb{R}^{|\Shat|}$. The corresponding statistic can be written as $T=\eta^TY$, with
$
\eta=X_{\Shat}(X_{\Shat}^TX_{\Shat})^{-1}a.
$
We also write $T_{\mathrm{obs}}=\eta^TY_{\mathrm{obs}}$. For Monte Carlo sampling under the null $a^T\beta^{\Shat}=c$, we use the restricted least-squares fitted mean
$$
\begin{aligned}
\hat\beta_{\Shat} &= (X_{\Shat}^TX_{\Shat})^{-1}X_{\Shat}^TY_{\mathrm{obs}},\\
\hat\beta^{0}_{\Shat}
&= \hat\beta_{\Shat}
-\frac{a^T\hat\beta_{\Shat}-c}{a^T(X_{\Shat}^TX_{\Shat})^{-1}a}
  (X_{\Shat}^TX_{\Shat})^{-1}a,\\
\mu_0 &= X_{\Shat}\hat\beta^{0}_{\Shat}.
\end{aligned}
$$
Here $\mu_0$ is the null mean used to generate the selective-model reference distribution.

When $\sigma$ is known, the selective-model reference law is
$
\mathcal L_{\mu_0,\sigma}\left\{\eta^TY\mid \mathcal A_{\AIC}(Y,X)=\Shat\right\}.
$
Because we do not condition on $P_{\eta^\perp}Y$, this conditional law is not reduced to the one-dimensional truncated normal pivot in Section~\ref{sec:constructCI}. We approximate it by accept--reject Monte Carlo: generate $Y_b\sim N(\mu_0,\sigma^2I_n)$, retain the draws for which AIC selects $\Shat$, and use the retained values of $\eta^TY_b$ to estimate the selective null distribution. Algorithm~\ref{alg:parametric_selective_bootstrap} summarizes this procedure.

When $\sigma$ is unknown, we remove the nuisance scale by conditioning on the radius
$
R_{\mathrm{obs}}=\|Y_{\mathrm{obs}}-\mu_0\|.
$
By spherical symmetry, conditional on this radius, the direction of $Y-\mu_0$ is uniformly distributed on the unit sphere $\mathbb S^{n-1}$. We therefore generate $Y_b=\mu_0+R_{\mathrm{obs}}v_b$, where $v_b$ is uniform on $\mathbb S^{n-1}$, and again retain only the draws for which AIC selects $\Shat$. This gives a Monte Carlo approximation to the fixed-radius selective-model null distribution; see Algorithm~\ref{alg:fixed_radius_selective_sampling}. More efficient samplers, such as hit-and-run Markov chain Monte Carlo (MCMC) or importance sampling, may be used when the accept--reject rate is small.

\begin{algorithm}[ht]
\caption{Selective $p$-value with known $\sigma^2$}
\label{alg:parametric_selective_bootstrap}
\nl \KwIn{$Y_{\mathrm{obs}}$, design matrix $X$, selection map $\mathcal A_{\AIC}(\cdot,X)$, null mean $\mu_0$, contrast vector $\eta$, noise level $\sigma^2$, number of proposals $B$}
\nl \KwOut{Accepted statistics $\mathcal T$, empirical selective null cdf $\widehat F_0$, and two-sided selective $p$-value $p_{\mathrm{sel}}$}

\nl Record the observed selection $\Shat\leftarrow\mathcal A_{\AIC}(Y_{\mathrm{obs}},X)$ and set $T_{\mathrm{obs}}\leftarrow\eta^TY_{\mathrm{obs}}$\;
\nl Initialize accepted index set $\mathcal I\leftarrow\emptyset$ and list $\mathcal T\leftarrow\emptyset$\;

\nl \For{$b=1,\ldots,B$}{
    \nl Generate $Y_b\sim N(\mu_0,\sigma^2I_n)$ independently\;
    \nl \If{$\mathcal A_{\AIC}(Y_b,X)=\Shat$}{
        \nl Compute $T_b\leftarrow\eta^TY_b$\;
        \nl Append $b$ to $\mathcal I$ and $T_b$ to $\mathcal T$\;
    }
}

\nl If $|\mathcal T|=0$, increase $B$ or use a more efficient conditional sampler\;
\nl Construct the empirical conditional cdf
$\widehat F_0(t)\leftarrow |\mathcal T|^{-1}\sum_{T\in\mathcal T}\mathbf 1\{T\le t\}$\;
\nl Compute $\pval (Y)_{\mathrm{sel}}\leftarrow 2\min\{\widehat F_0(T_{\mathrm{obs}}),1-\widehat F_0(T_{\mathrm{obs}})\}$\;
\nl \Return $\mathcal T$, $\widehat F_0$, and $\pval(Y)_{\mathrm{sel}}$\;
\end{algorithm}

\begin{algorithm}[ht]
\caption{Selective $p$-value with unknown $\sigma^2$}
\label{alg:fixed_radius_selective_sampling}
\nl \KwIn{$Y_{\mathrm{obs}}$, design matrix $X$, selection map $\mathcal A_{\AIC}(\cdot,X)$, null mean $\mu_0$, contrast vector $\eta$, number of proposals $B$}
\nl \KwOut{Accepted statistics $\mathcal T$, empirical selective null cdf $\widehat F_0$, and two-sided selective $p$-value $p_{\mathrm{sel}}$}

\nl Record the observed selection $\Shat\leftarrow\mathcal A_{\AIC}(Y_{\mathrm{obs}},X)$ and set $T_{\mathrm{obs}}\leftarrow\eta^TY_{\mathrm{obs}}$\;
\nl Compute the observed radius $R_{\mathrm{obs}}\leftarrow\|Y_{\mathrm{obs}}-\mu_0\|$\;
\nl Initialize accepted index set $\mathcal I\leftarrow\emptyset$ and list $\mathcal T\leftarrow\emptyset$\;

\nl \For{$b=1,\ldots,B$}{
    \nl Generate $g_b\sim N(0,I_n)$ and set $v_b\leftarrow g_b/\|g_b\|$\;
    \nl Propose $Y_b\leftarrow\mu_0+R_{\mathrm{obs}}v_b$\;
    \nl \If{$\mathcal A_{\AIC}(Y_b,X)=\Shat$}{
        \nl Compute $T_b\leftarrow\eta^TY_b$\;
        \nl Append $b$ to $\mathcal I$ and $T_b$ to $\mathcal T$\;
    }
}

\nl If $|\mathcal T|=0$, increase $B$ or use a more efficient conditional sampler\;
\nl Construct the empirical conditional cdf
$\widehat F_0(t)\leftarrow |\mathcal T|^{-1}\sum_{T\in\mathcal T}\mathbf 1\{T\le t\}$\;
\nl Compute $\pval(Y)_{\mathrm{sel}}\leftarrow 2\min\{\widehat F_0(T_{\mathrm{obs}}),1-\widehat F_0(T_{\mathrm{obs}})\}$\;
\nl \Return $\mathcal T$, $\widehat F_0$, and $\pval(Y)_{\mathrm{sel}}$\;
\end{algorithm}

Compared with the saturated-model test, the selective-model test conditions on less information and can therefore be more powerful. The trade-off is that it evaluates the null distribution under $H_0^{\mathrm{sel}}$, rather than under the unrestricted saturated-model null $H_0^{\mathrm{sat}}$. The selective-model procedure is therefore useful for testing, especially when $\sigma$ is unknown, but it is computationally more demanding because inference requires sampling conditional on the reselection event.

\begin{figure}[H]
    \centering
    \includegraphics[width=0.99\linewidth]{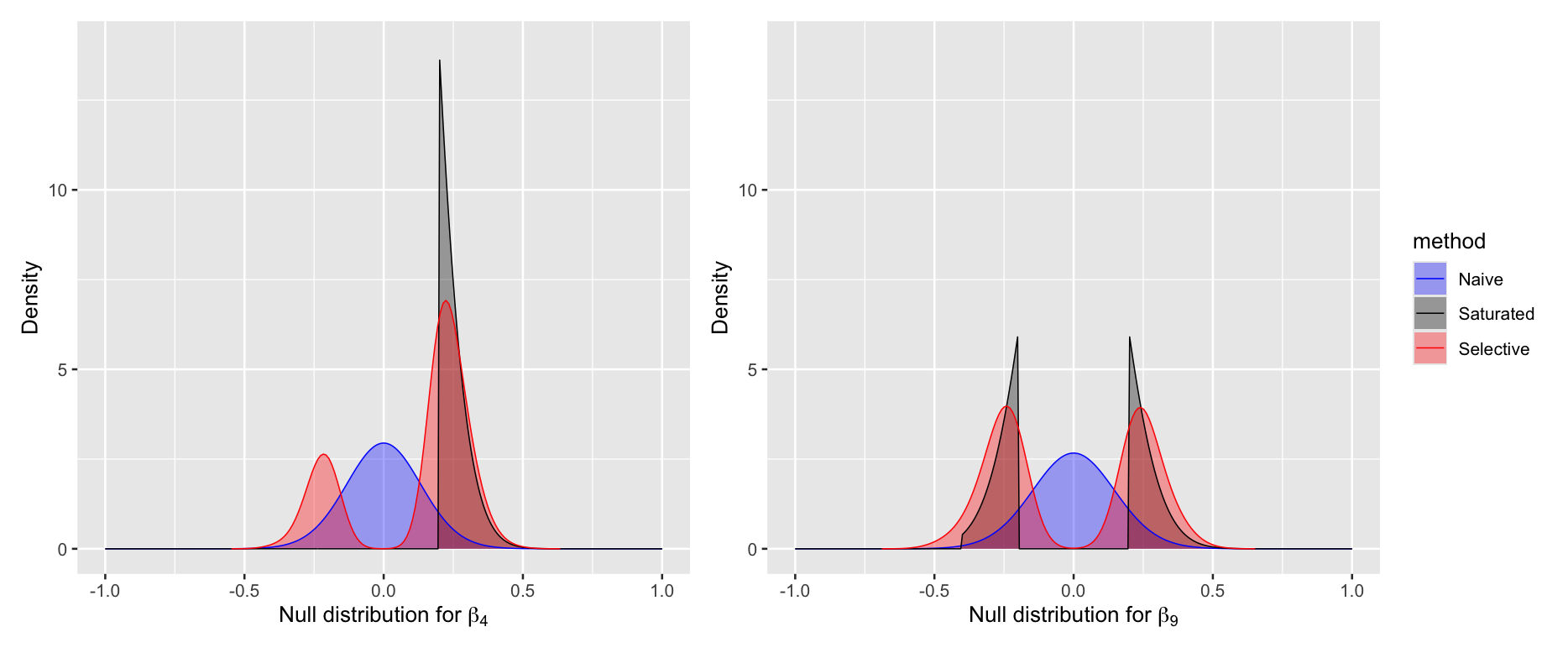}
    \caption{Comparison of the pivotal-quantity distributions at true nulls for the naive, saturated, and selective tests. The simulation setting follows the motivating example in Section~\ref{sec:moti}, with $p=10$, $n=30$, and $\beta=(1,2,3,0,\ldots,0)$.}
    \label{fig:pivot}
\end{figure}

Figure~\ref{fig:pivot} compares the distributions of pivotal quantities for two falsely selected coefficients, $\beta_4$ and $\beta_9$, both of which have true value zero, in the motivating example with known variance. We consider the naive, saturated-model, and selective-model tests. The naive test follows the classical procedure and yields a Gaussian pivotal distribution. In contrast, the saturated and selective-model pivots are shaped by conditioning on the variable-selection event. For the selective-model test, the selection effect further distorts the distribution, turning a high-density region of the naive pivot into a low-density ``valley.'' This effect has no closed-form expression, so we approximate the selective-model null distribution using the Monte Carlo procedure described above.

Section~\ref{sec:simu} compares the plug-in saturated-model approach and the selective-model selective test in finite samples. In the small-sample unknown-variance setting, the plug-in saturated-model test can exhibit Type~I error inflation, whereas the selective-model test is closer to the nominal level. As the sample size increases, the plug-in inflation decreases. The main practical limitation of the selective-model test is computation: the accept--reject step can be expensive when the reselection probability
$
\Pr\{\mathcal A_{\AIC}(Y,X)=\Shat\mid H_0^{\mathrm{sel}}\}
$
is small. For this reason, inverting the selective-model test to construct confidence intervals can be costly.

\section{Simulation}\label{sec:simu}

We conduct numerical experiments to evaluate the finite-sample behavior of the proposed post-selection inference procedures. First, we examine whether the corrected confidence intervals attain nominal coverage after conditioning on the AIC selection event. Second, we compare the post-AIC Type~I error and power of the naive, saturated, and selective-model tests in a small-sample setting. The simulations show that the proposed adjustments provide valid post-AIC confidence intervals and $p$-values under known variance and substantially improve calibration when the variance is unknown.

\subsection{Coverage of post-AIC confidence intervals}
\label{sec:simu_ci}

We first examine confidence-interval coverage after AIC selection. Empirical coverage is computed from 5{,}000 simulations. The data-generating mechanism is the same as in Section~\ref{sec:moti}; in these simulations, AIC selects either the true model or an overfitted model. The confidence level is $1-\alpha=0.95$, with $\alpha=0.05$. For each replication, we construct confidence intervals at ten new $x$ points and compare the classical uncorrected intervals with the proposed post-AIC corrected intervals.

\begin{figure}[htbp]
\centering
\minipage{\textwidth}
\includegraphics[width=\linewidth]{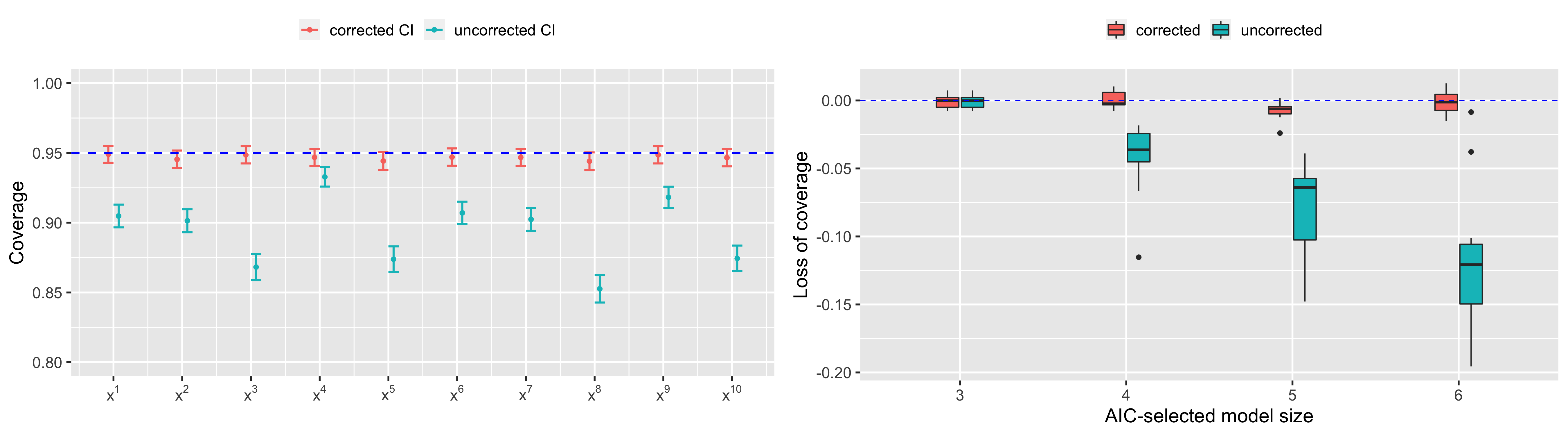}
\endminipage\hfill
\caption{Coverage of uncorrected and corrected confidence intervals when $\sigma$ is known. Left: coverage probabilities, with 95\% Monte Carlo error bars, at each new data point. Right: boxplots of empirical coverage minus 0.95, grouped by selected model size. Negative values indicate undercoverage.}
\label{fig:simu_knownsigma}
\end{figure}

We begin with the setting $\sigma=1$ known. Figure~\ref{fig:simu_knownsigma} reports the empirical coverage of the uncorrected and corrected confidence intervals. The left panel shows that the uncorrected intervals exhibit undercoverage at all ten new $x$ points. In contrast, the proposed correction brings empirical coverage close to the nominal level. Averaged across the ten points, Table~\ref{tab:cover_loss} shows that coverage increases from 0.894 for the uncorrected known-$\sigma$ interval to 0.947 after correction.

The right panel of Figure~\ref{fig:simu_knownsigma} breaks down coverage by selected model size. We show only selected model sizes that occur in more than 5\% of the 5{,}000 simulations. In this experiment, whenever the selected model size is three, the true model is selected. In that case, the classical intervals have little coverage loss, with or without correction. Coverage loss becomes substantial when AIC selects a strictly overfitted model and becomes more severe as the selected model grows. This pattern explains the overall undercoverage of the uncorrected intervals. After correction, coverage is close to nominal across selected model sizes, suggesting that the loss is primarily driven by ignoring the AIC selection event.

To illustrate how the selection event changes the conditional distribution, Table~\ref{tab:intervals} reports the excluded region of the conditional truncated normal distribution in four representative examples, one for each selected model size in $\{3,4,5,6\}$. The table also compares the confidence intervals before and after correction. The true value of the estimand is 9.53, which is the mean parameter $\mu$ in the truncated normal distribution. For selected model sizes 3 and 4, the excluded region has little probability mass under the corresponding normal distribution, so the correction has almost no visible effect on the interval. For selected model size 6, however, the excluded region overlaps with the high-density region around 9.53, leading to a noticeably wider corrected interval. Thus, the difference between corrected and uncorrected intervals is driven by how much probability mass the selection event removes from the original pivot distribution.

\begin{table}[ht]
    \centering
    \begin{tabular}{cccc}
    \toprule
        size & excluded region & \multicolumn{2}{c}{confidence interval} \\
        & & uncorrected & corrected  \\
        \hline
        3 & $(-7.2, -4.8)\cup (-0.4, 1.5)\cup (12.4, 14.2)$ & $(9.38, 10.77)$ & $(9.38, 10.77)$ \\
        4 & $(-9.7,-6.8)\cup (1.4, 8.6)\cup (11.2,13.2)$ & $(9.11,10.51)$ & $(9.10, 10.51)$\\
        5 & $(-17.8, -13.8)\cup (-1.9, 0.8)\cup (9.8,13.0)\cup (14.2, 16.6)$ & $(8.60,10.18)$ & $(8.61,10.99)$\\
        6 & $(-19.5,-15.6)\cup (-2.5,-0.2)\cup (9.4, 14.6)$ & $(8.36,9.95)$ & $(8.41,11.74)$\\
        \bottomrule
    \end{tabular}
    \caption{Examples of excluded regions in the conditional truncated normal distribution and the corresponding confidence intervals before and after correction, for different AIC-selected model sizes.}
    \label{tab:intervals}
\end{table}

We next consider the setting in which $\sigma$ is unknown. We use a plug-in approach: estimate $\sigma$ by $\tilde\sigma$ and construct corrected confidence intervals using $\tilde\sigma$ in place of $\sigma$. We consider two mean-squared-error estimators,
\[
\tilde{\sigma}^2 = \RSS(S)/(n-|S|-1).
\]
The first uses the full model for $S$ and is denoted \texttt{mse\_full}. The second uses the selected model $\widehat S$ for $S$ and is denoted \texttt{mse\_aic}. The full-model estimator is consistent when the true model is nested in the full model, whereas the selective-model estimator is the classical choice when the selected model is treated as fixed but may be biased after selection.

Table~\ref{tab:cover_loss} summarizes empirical coverage. Even when $\sigma$ is unknown, the proposed correction substantially mitigates the undercoverage of the classical intervals. Using the same selective-model estimate of $\sigma$, the average coverage improves from 0.883 for the uncorrected interval to 0.936 for the corrected interval with \texttt{mse\_aic}. A small amount of coverage loss remains relative to the nominal level. The \texttt{mse\_full} estimator performs better in this experiment, with average coverage 0.944. Across the 5{,}000 replications, \texttt{mse\_aic} slightly underestimates $\sigma$, with mean 0.961, whereas \texttt{mse\_full} is closer to the truth, with mean 0.996. These results show that post-AIC correction largely restores coverage, but variance estimation remains important in small samples.

\begin{table}[ht]
    \centering
\begin{tabular}{cccccc}
\toprule
& \multicolumn{2}{c}{Uncorrected CI} & \multicolumn{3}{c}{Corrected CI} \\ 
\cmidrule(lr){2-3} \cmidrule(lr){4-6} 
 & $C(\eta(x), S, \alpha)$ & $C_{\sigma}(\eta(x), S, \alpha)$ & known $\sigma$ & \texttt{mse\_aic} & \texttt{mse\_full} \\
\cmidrule(lr){2-3} \cmidrule(lr){4-6} 
Coverage  & 0.883 & 0.894 & 0.947 & 0.936 & 0.944 \\
Relative loss (\%)  & 6.74 & 5.64 & 0.32 & 1.47 & 0.63 \\
\midrule
& $C(\eta(x^1), S, \alpha)$ & $C_{\sigma}(\eta(x^1), S, \alpha)$ & known $\sigma$ & \texttt{mse\_aic} & \texttt{mse\_full} \\
\cmidrule(lr){2-3} \cmidrule(lr){4-6}
Coverage & 0.892 & 0.905 & 0.949 & 0.938 & 0.945 \\
Standard error & 0.004 & 0.004 & 0.003 & 0.003 & 0.003 \\
Relative loss (\%) & 6.15 & 4.76 & 0.11 & 1.22 & 0.57 \\
\bottomrule
\end{tabular}
    \caption{Empirical coverage of confidence intervals with confidence level $1-\alpha=0.95$. The upper panel reports average coverage across ten new $x$ points, and the lower panel reports coverage for $x^1$. Relative coverage loss is $1-\{(\text{empirical coverage})/0.95\}$, set to zero if negative. The first two columns are classical uncorrected intervals. The last three columns are corrected intervals using the proposed method with known $\sigma$, selective-model variance estimation, and full-model variance estimation, respectively.}
    \label{tab:cover_loss}
\end{table}

The remaining coverage loss under \texttt{mse\_aic} motivates the finite-sample Type~I error analysis in the next subsection. We now examine Type~I error after conditioning on the AIC selection event under both known and unknown noise levels.

\subsection{Finite-sample Type~I error and power after selecting \texorpdfstring{$\beta_2$}{beta 2}}
\label{sec:simu_typeI}

We next compare the Type~I error and power of three post-selection tests. We simulate a linear regression model with true coefficient vector
\[
\boldsymbol{\beta} = (1,\beta_2,2,0,0.5).
\]
For each value of $\beta_2$, we apply AIC best subset selection and test $H_0:\beta_2=0$, conditional on $\beta_2$ being selected. We repeatedly generate datasets until we obtain 1{,}000 samples in which $\beta_2$ is included in the selected model, and we compute rejection rates from these selected samples.

We compare three methods. The naive method ignores selection and uses the usual $z$-test when $\sigma$ is known and the usual $t$-test when $\sigma$ is unknown. The saturated-model test adjusts for the AIC selection event and derives the truncation region for the null distribution; when $\sigma$ is unknown, it uses a plug-in estimate of $\sigma^2$ from the selected model. The selective test conditions on less information and computes each $p$-value using 1{,}000 Monte Carlo samples generated under the selection event. We focus on the small-sample setting $n=30$ and $\sigma=1$, where the finite-sample effects of variance estimation are most visible.

\begin{figure}[htbp]
  \centering
  \includegraphics[width=0.9\textwidth]{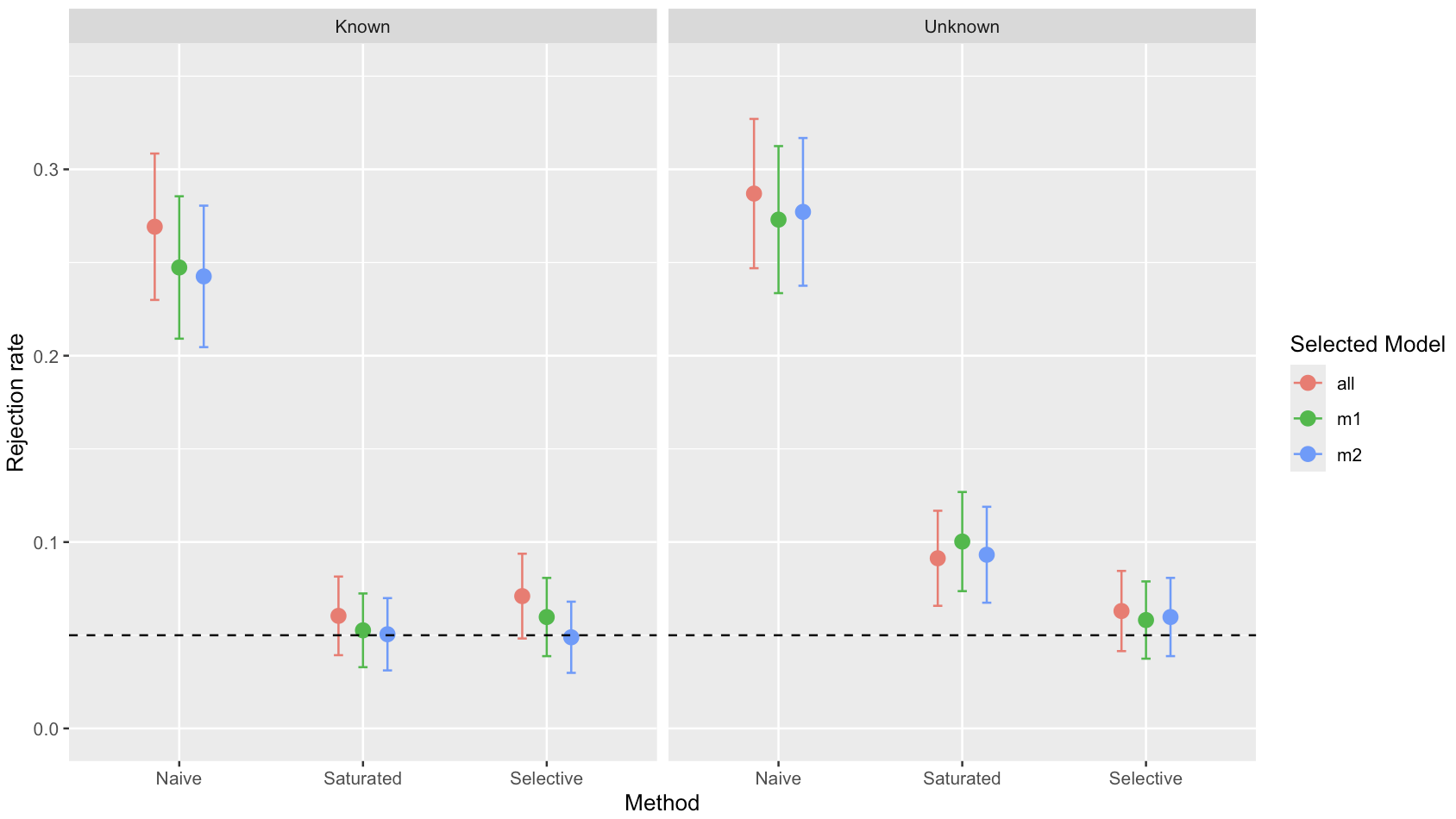} 
  \caption{Type~I error rate at $\beta_2=0$ for the naive, saturated, and selective tests under known and unknown $\sigma$. Dashed line represents the nominal rejection rate 0.05. Results are shown marginally over all selected models containing $\beta_2$ and conditionally on two selected models, $m_1$ and $m_2$. In the unknown-variance saturated test, $\sigma^2$ is estimated from the selected model; for $m_2$, this coincides with the full-model MSE estimator because $m_2$ is the full model.}
  \label{fig:conditional}
\end{figure}

Figure~\ref{fig:conditional} reports the Type~I error rate at the null, $\beta_2=0$, under both known and unknown variance. The figure includes the marginal rejection rate over all selected models containing $\beta_2$ and conditional rejection rates for two selected models. The first model, $m_1$, is the event that $(\beta_1,\beta_2,\beta_3,\beta_5)$ are selected; under the null, this is the true model. The second model, $m_2$, is the event that $(\beta_1,\beta_2,\beta_3,\beta_4,\beta_5)$ are selected; in this case, the selected model is overfitted.

For the saturated-model test with unknown variance, the plug-in estimate of $\sigma^2$ is computed from the selected model. This is the selective-model estimator \texttt{mse\_aic} used in the confidence-interval experiment. In Figure~\ref{fig:conditional}, however, the conditional case $m_2=(\beta_1,\beta_2,\beta_3,\beta_4,\beta_5)$ is the full model. Hence, conditional on $m_2$, the selective-model and full-model estimators are identical. We therefore do not add a separate \texttt{mse\_full} saturated-test curve in Figure~\ref{fig:conditional}. The $m_2$ conditional result already represents that case.

In Figure~\ref{fig:conditional}, the naive test substantially overrejects, with Type~I error around 28\%. The saturated test with plug-in variance estimation also overrejects, with Type~I error around 10\%. This inflation occurs because the plug-in estimate of $\sigma^2$ does not fully account for the additional randomness introduced by model selection and variance estimation. In contrast, the selective test is much closer to the nominal 5\% level, with Type~I error around 6\%. Thus, in this small-sample setting, the selective test gives better finite-sample calibration when $\sigma$ is unknown. When $\sigma$ is known, both adjusted methods control Type~I error close to the nominal level, while the naive test remains severely anti-conservative. This confirms that Type~I error inflation is mainly caused by ignoring the selection event and that conditioning on the AIC event restores finite-sample validity when the nuisance variance is known.

\begin{figure}[htbp]
  \centering
  \includegraphics[width=0.99\textwidth]{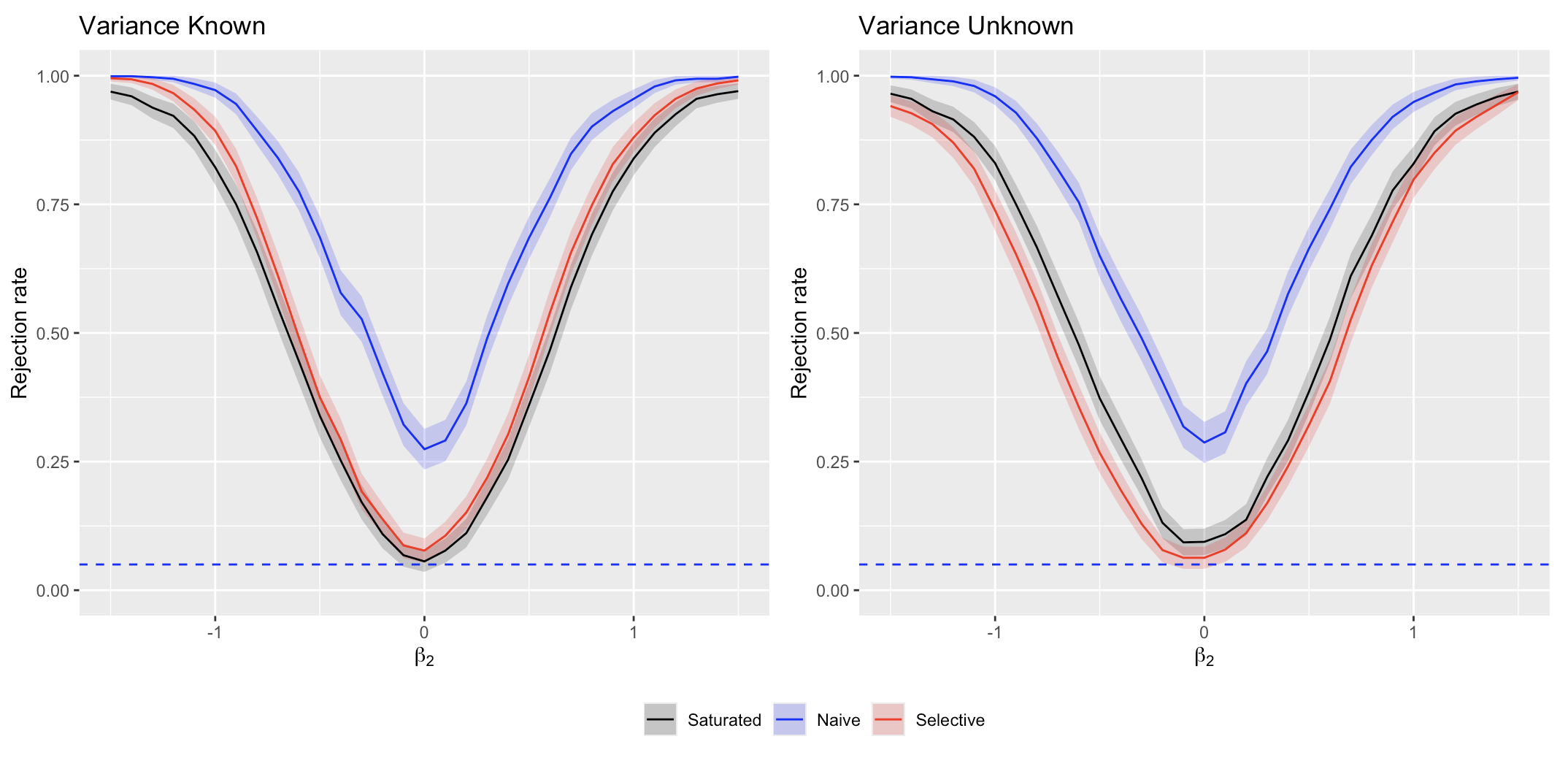} 
  \caption{Marginal rejection rate comparison among the naive, saturated, and selective tests for post-AIC inference. Dashed line represents the nominal rejection rate 0.05. The rejection rate is plotted as a function of $\beta_2$, conditional on $\beta_2$ being selected by AIC.}
  \label{fig:comparison}
\end{figure}

Next, we compare the full rejection-rate curves as $\beta_2$ varies. Figure~\ref{fig:comparison} shows the marginal rejection rate of the naive, saturated, and selective tests as a function of $\beta_2$. In the known-variance setting, the naive test has the highest power, but it fails to control Type~I error under the null. Among the valid adjusted methods, the selective test is slightly more powerful than the saturated test as the signal strength increases. This agrees with the theoretical expectation that conditioning on less information can retain more information for inference. In this experiment, however, the power gain is modest.

In the unknown-variance setting, the naive test again fails to control Type~I error. The saturated test also exhibits noticeable Type~I error inflation because of plug-in variance estimation. The selective test remains closer to nominal, although it is computationally more expensive: each selective $p$-value requires Monte Carlo sampling under the selection event. This computational cost makes the selective test less convenient for constructing confidence intervals and for extending the method to more complex estimands.

Overall, the simulations show that the proposed post-selection adjustments substantially correct finite-sample inferential distortions caused by AIC selection. The confidence-interval experiment demonstrates large improvements in coverage after correction, while the Type~I error experiment shows that valid post-selection testing requires accounting for the selection event. The remaining discrepancies in the unknown-variance case show that the saturated method with plug-in variance estimation can still inflate Type~I error. The selective approach is computationally heavier but gives the best finite-sample calibration in this setting.

\section{Real data application} \label{sec:application}
We apply the post-AIC inference procedure to the U.S. consumption dataset used by \citet{hyndman2018forecasting}. In this application, four predictors--personal disposable income (Income), industrial production (Production), savings (Savings), and the unemployment rate (Unemployment)--are used to predict percentage changes in quarterly personal consumption expenditure from 1970 to 2016. The sample size is $n=198$. The dataset is available in the R package \texttt{fpp3} under the name \texttt{us\_change}.

In Section 7.5 of \citet{hyndman2018forecasting}, AIC selected the full model. A notable feature of the example is the separation in AIC values between models containing both Income and Savings and models missing at least one of them. The AIC values of the former models range from $-456.6$ to $-435.7$, whereas the smallest AIC value among the latter models is $-262.3$. Thus, excluding Income or Savings sharply increases AIC, while excluding either Unemployment or Production from the full model has little effect. This suggests that Income and Savings are more important predictors than Production and Unemployment. However, at significance level $\alpha=0.05$, the conventional 95\% confidence intervals for Income, Savings, and Production all exclude zero (Table~\ref{tab:realdtci}), providing no such separation between Income/Savings and Production.

When the full model is selected, as in this application, Lemma~\ref{lem:aiceffect_smallS} suggests that variables with weak evidence can appear significant if classical inference is applied after AIC selection without accounting for the randomness of selection. We apply the proposed post-AIC correction and report 95\% post-AIC confidence intervals and $p$-values in Table~\ref{tab:realdtci}. The noise level is estimated using the strategies described in Section~\ref{sec:simu}; here \texttt{mse\_aic} coincides with \texttt{mse\_full} because the selected model is the full model. After correction, only Income and Savings remain significant. Production is no longer significant because its corrected 95\% confidence interval contains zero. The same qualitative conclusion is obtained from the saturated plug-in and selective tests. Thus, the proposed post-best-subset AIC inference leads to a different conclusion from conventional inference and better matches the empirical ranking suggested by the AIC values.

\begin{table}[ht]
    \centering
    \begin{tabular}{lccc}
    \toprule
        & uncorrected & corrected (\texttt{mse\_full}) & selective \\
    \midrule
    \multicolumn{4}{l}{\textit{95\% confidence intervals}}\\
    Income & $(0.6615, 0.8197)$ & $(0.6620, 0.8245)$ & -- \\
    Production & $(0.0015, 0.0928)$ & $(-0.0109, 0.1148)$ & -- \\
    Savings & $(-0.0587, -0.0471)$ & $(-0.0593, -0.0472)$ & -- \\
    Unemployment & $(-0.3631, 0.0137)$ & $(-0.4564, 0.0623)$ & -- \\
    \midrule
    \multicolumn{4}{l}{\textit{$p$-values}}\\
    Income & $<0.0001$ & $<0.0001$ & $<0.0001$ \\
    Production & $0.0428$ & $0.2114$ & $0.104$\\
    Savings & $<0.0001$ & $<0.0001$ & $<0.0001$\\
    Unemployment & $0.0689$ & $0.3853$ & $0.300$\\
    \bottomrule
    \end{tabular}
    \caption{Confidence intervals and $p$-values for the U.S. consumption application. The uncorrected columns report the conventional 95\% confidence intervals and $p$-values from the AIC-selected model without post-selection correction. The corrected \texttt{mse\_full} column reports post-AIC corrected 95\% confidence intervals and $p$-values with $\sigma$ estimated from the full model; because AIC selects the full model, this is the same as \texttt{mse\_aic}. The selective column reports Monte Carlo selective-model $p$-values; selective confidence intervals were not computed because of the computational cost of inverting the Monte Carlo test.}
    \label{tab:realdtci}
\end{table}

\section{Discussion}\label{sec:discuss}
Classical confidence intervals after best subset selection are widely implemented in statistical software and are routinely used in scientific applications to assess significance. In this paper, we construct post-selection corrected confidence intervals for linear models selected by AIC. Conditional on the best subset AIC selection event, a linear estimator is no longer normally distributed; instead, its sampling distribution is a truncated normal distribution over a finite union of intervals. This characterization yields corrected confidence intervals for a broad class of linear combinations of regression coefficients. The AIC selection event has an interval-set representation, and the resulting post-AIC confidence intervals do not require the extra conditioning often used in other selective-inference procedures. In fact, the implementation can usually be simplified to less-than-exact conditioning without sacrificing validity.

The correction extends directly to best subset selection under other information-based criteria, because many such criteria differ from AIC mainly in the strength of the model-size penalty. For example, the proposed inference is readily applicable to the Bayesian information criterion (BIC) and the corrected AIC (AICc) by adjusting $\omega(S)$ in~\eqref{eq:ab.set}. In particular, $\omega(S)$ is $\exp\{\log n(|\Shat|-|S|)/n\}$ for BIC and $\exp\{2(|\Shat|/(n-|\Shat|-1)-|S|/(n-|S|-1))\}$ for AICc. The accompanying implementation, available at \url{https://github.com/RiceOwlXinTan/best-subset-regression-inference}, includes AIC, BIC, and AICc. Mallows' $C_p$ also falls within the scope of the methodology because conditioning on the selected model again yields a sequence of quadratic constraints on the estimator, although the correction requires more than a change to $\omega(S)$.

We also study post-selection inference with unknown noise level and compare the common plug-in approach with a selective-model Monte Carlo test. In small samples, the plug-in approach under the saturated-model framework can still inflate Type~I error, while the selective-model test maintains near-nominal Type~I error in our experiments. Future work could compare post-best-subset inference with post-sequential-selection inference to better understand how different selection mechanisms affect the pivotal quantity.

A further direction is scalable post-AIC confidence inference for high-dimensional models. Finding the best model under information-based objectives requires evaluating exponentially many candidates, which is computationally infeasible in high dimensions. Lemma~\ref{lem:aiceffect_largeS} reduces the computational burden of deriving the conditional sampling distribution, but it does not change the complexity of finding the selected model. Recent advances have accelerated best subset selection through mixed-integer optimization~\citep{bertsimas2016best} and polynomial-time algorithms~\citep{zhu2020polynomial}. Building on these developments could lead to scalable characterizations of the selection event and faster construction of post-selection confidence intervals for large datasets.

\bibliographystyle{apalike}
\bibliography{postAIC}

\end{document}